\documentstyle[12pt,worldsci]{article}
\def\sss{\scriptscriptstyle}
\def\lft{{\sss L}}
\def\rht{{\sss R}}
\def\subs{{\sss S}}
\def\subt{{\sss T}}
\begin{document}

\noindent
UdeM-LPN-TH-93-160 \\
McGill-93/29 \\
September, 1993

\title{{\bf SINGLE LEPTOQUARK PRODUCTION}\footnote{Invited talk given by
G.B. at the ``Workshop on Physics and Experiments at Linear $e^+e^-$
Colliders'', Waikoloa, Hawaii, April 26-30, 1993.}}

\author{G. B\'ELANGER${}^a$, D. LONDON${}^a$ and H. NADEAU${}^b$
\vskip1truemm
{\em ${}^a$ Laboratoire de Physique Nucl\'eaire, Universit\'e de
Montr\'eal,\\
C.P. 6128, Succ. A, Montr\'eal, Qu\'ebec, Canada H3C 3J7}
\vskip2truemm
{\em ${}^b$ Physics Department, McGill University, \\
3600 University St., Montr\'eal, Qu\'ebec, CANADA, H3A 2T8}}

\maketitle
\setlength{\baselineskip}{2.6ex}

\begin{center}
\parbox{13.0cm}
{\begin{center} ABSTRACT \end{center}
{\small \hspace*{0.3cm}
\noindent
The single production of leptoquarks in $e^+e^-$, $e\gamma$ and $\gamma
\gamma$ linear colliders is discussed. We show that these new particles
could be seen in such machines even if their mass is close to the kinematic
limit.
}}

\end{center}

Leptoquarks are predicted in many extensions of the standard model.
Although there is no compelling argument that, if they exist, their mass
will be low enough that they could be produced in the next generation of
colliders, the existence of leptoquarks would be such a striking
breakthrough in the search for the physics beyond the standard model that
it is important to look for them. Linear colliders would offer a clean
environment to do so. The obvious mode to produce leptoquarks is via pair
production. While this mode is very interesting and has sizeable
cross-sections, it has the drawback that only leptoquarks of masses up to
half the center-of-mass energy can be probed. Such leptoquarks will soon be
severely constrained by HERA.$^{1}$ On the other hand, the single
production mechanism allows one to probe twice as large a mass range.

In order to avoid any theoretical bias we will consider all possible
leptoquarks and classify them according to their $SU(3)\times SU(2)\times
U(1)$ quantum numbers. The most general model-independent Lagrangian$^2$
that describes the coupling of scalar leptoquarks to fermions can be
written as
\begin{eqnarray}
{\cal L} = ~ g_{1\lft} \, {\overline{q^c_\lft}}i\tau_2 l_\lft \, S_1
+ g_{3\lft} \, {\overline{q_\lft^c}}i\tau_2 \tau^i l_\lft \, S_3^i
+ h_{2\rht} \, {\overline q}_\lft i \tau_2 e_\rht \, R_2^\prime~+
{}~g_{1\rht} \, {\overline{u_\rht^c}}e_\rht \, S_1^\prime  \nonumber\\
+ {\tilde g}_{1\rht} \, {\overline{d_\rht^c}}e_\rht \, {\tilde S}_1
+ h_{2\rht} \, {\overline u}_\rht l_\lft \, R_2
+ {\tilde h}_{2\rht} \, {\overline d}_\rht l_\lft \, {\tilde R}_2~.
\end{eqnarray}
Eight different types of interactions of leptoquarks and fermions are
described here, namely the right- and left-handed couplings to either $e u,
e d, e \bar d$ or $e \bar u$ (the neutrino couplings are irrelevant),
which could make a general analysis of single production rather messy.
However, the situation is simpler than it appears since the unpolarized
cross-sections for the processes we will study are the same for both right-
and left-handed leptoquarks. Basically there will be only four types of
leptoquarks to consider, those of charge $Q_\subs=-1/3,-2/3,-4/3$ or
$-5/3$. In the event that leptoquarks were found, by using polarization it
would be a simple task to distinguish the right- and left-handed couplings.
Besides the charge and the mass of the leptoquark, the other free parameter
is the strength of the coupling, defined as $k=g^2/(4\pi \alpha_{em})$
where $g$ can be any of the coupling constants defined in Eq.\ 1. We will
refer to $k=1$ as a coupling of electromagnetic strength.

The possibility of leptoquarks coupling to different generations should not
be ignored. Here we will assume generation-diagonal leptoquarks since
non-generation diagonal ones tend to induce rare decays and are more
strongly constrained, typically to the multi-TeV scale.$^{3}$ We will also
assume chiral leptoquarks, which are less severely constrained than
non-chiral ones -- the limits are around $M_\subs/g> 300$ GeV (10 TeV) for
chiral (non-chiral) couplings.$^4$  While fully realizing that these
constraints are model dependent and could be evaded, we make these
assumptions to emphasize the role of a linear collider in searching for
leptoquarks. The important point is that there is still a lot of room for
linear colliders to improve on the limits from both indirect and direct
searches. This is particularly true for the single production mechanism
which probes the region inaccessible to direct searches at
HERA.\footnote{By indirect searches HERA will be able to significantly
improve on those limits.$^5$} The present limit from this collider is
$M_\subs >100-200$ GeV for couplings of electromagnetic
strength.$^1$

\begin{figure}
\vspace{3.2cm}
\caption{Diagrams giving the dominant contribution to
$e^+e^-\rightarrow Se^+q$}
\end{figure}

First consider the single production of leptoquarks in the process $e^+ e^-
\rightarrow S e^+ q$. A large number of diagrams contribute to this
process. However, a quick inspection allows for the selection of the
dominant diagrams, namely those in which the amplitude diverges in the
forward or backward direction. These singularities are regulated by the
mass of the lepton or quark involved, giving rise to logarithmic
enhancements of the cross section by factors of $\log(s/m^2)$, hereafter
called `large logs'. In Fig.\ 1, only the dominant diagrams are shown --
the first two contain one large log while the last contains two such large
logs. Since the dominant term for the cross section is proportional to
$(Q_\subs+1)^2$, we therefore expect $\sigma_{Q_\subs=-1/3} \approx
\sigma_{Q_\subs=-5/3} \approx 4\sigma_{Q_\subs=-4/3} \approx 4
\sigma_{Q_\subs=-2/3}$. This is confirmed numerically$^6$ as shown in
Fig.\ 2. Note that for this process the forward divergence can be
alternatively regulated with a $p_\subt$ cut (as was done in the first
calculation of this process$^7$). The advantage of using a mass regulator
is that a much larger fraction of the events are kept.
In Table
1 we give the results for discovery limits for leptoquarks with couplings
of electromagnetic strength for a 500 GeV and 1 TeV collider.$^6$ The
criterion for discovery was fixed at 25 events. We can conclude that
leptoquarks can be discovered up to essentially the kinematic limit.
There is also the
possibility of searching for second or third generation leptoquarks through
the non-dominant diagrams with $s$-channel $\gamma/Z$ exchange.
Unfortunately, the cross-section from these diagrams is just too small
($\sigma \approx 10^{-3} fb$).

\begin{figure}
\vspace{8.5cm}
\caption{Cross-sections for single leptoquark production in $e^+e^-$ at (a)
$\protect\sqrt{s}=500$ GeV (b) $\protect\sqrt{s}=1$ TeV for the four
different leptoquark charges and $k=1$}
\end{figure}

The three diagrams that contribute to $e \gamma\rightarrow q S$ are those
shown in Fig.\ 1 for the $e^+ e^-$ process. The only difference is that the
photon is obtained by back-scattering laser light off the lepton beam.
Again the last diagram gives the dominant contribution due to the large log
coming from the collinear singularity in the t-channel regulated by the
mass of the quark.
The numerical results are given in Ref.\ 8, and the
conclusion is that leptoquarks will be observable up to the kinematic limit
for $k=1$. More weakly coupled leptoquarks are also observable. For
example, with the requirement of 25 events and a luminosity of $10 fb^{-1}$
at $\sqrt s=500$ GeV, leptoquarks of $M_\subs=400$ GeV are detectable for
$k\approx 0.01,0.02,0.03$ and $0.08$ for $Q_\subs=-5/3,-1/3,-4/3$ and
$-2/3$ respectively. Even heavier leptoquarks would be observable if we
consider the virtual production of a leptoquark. In this case, however, we
face an important standard model background from $e\gamma \rightarrow e q
\bar{q}$.

\vskip.6cm
\baselineskip=10pt
{\bf TABLE 1.} Maximum leptoquark mass observable in single production.
\vglue.1in
\baselineskip=12pt
\begin{tabular}{|c|c|c|c|}\hline
Process& $Q_\subs$& ${(M_\subs)}_{max}$ (GeV)&${(M_\subs)}_{max}$ (GeV)\\
&&$\sqrt{s}=500$ GeV&$\sqrt{s}=1$ TeV\\
&&${\cal L}=10 fb^{-1}$&${\cal L}=60 fb^{-1}$\\
\hline\hline
$e^+e^-\rightarrow e^+ qS$& $-$1/3, $-$5/3& 475 & 960\\
& $-$2/3, $-$4/3& 420 & 870\\
\hline
$\gamma\gamma\rightarrow e^+ qS$& $-$1/3, $-$5/3& 480 & 970\\
& $-$2/3, $-$4/3& 425 & 920\\
 \hline
\end{tabular}
\vskip.6cm

For the process $\gamma\gamma\rightarrow e^+ q S$, there are
twelve diagrams that contribute. With the same method of looking for large
logs it is not hard to convince oneself that only two dominate, the
lepton-quark fusion diagram of Fig.\ 3, and its partner under
symmetrization of the photons. Such diagrams will contain two large logs.
In Table 1 we give the results for the upper limit
of observability of leptoquarks again using the criterion of 25 events.$^6$
We see that in this process also, leptoquarks of electromagnetic strength
can be observed almost up to the kinematic limit. An important point to
remember, however, is that in a $\gamma \gamma$ collider, one will not be
able to achieve the same center of mass energy as the parent $e^+ e^-$
collider since the back-scattered photons obtained from shooting a laser at
the lepton beam would have a certain energy spread. Typically we expect to
have at most 80\% of the energy of the comparable electron machine.

An important feature of the $\gamma \gamma$ process is that one can produce
in the same way all generations of leptoquarks, opening up the possibility
of producing leptoquarks that couple mainly to second and third generation
fermions. One expects, however, that the large logs will not be as
important for heavier fermions. By a simple estimate of the ratio of the
logarithmic factors involved, we expect the second generation leptoquark
production cross-section to be suppressed by a factor 2 or 3 relative to
first generation leptoquarks, while that of a $\tau b$ leptoquark should be
suppressed by a factor of 7, and that of a $\tau t$ leptoquark by a factor
of 25 for $m_t=150$ GeV. The numerical calculation of the cross-sections
confirms these naive estimates and full details and the corresponding
graphs are given in Ref.\ 6.

In conclusion, leptoquarks which couple mainly to the first generation can
be singly produced in $e^+e^-,e\gamma$ or $\gamma\gamma$ collisions almost
up to the kinematic limit if their coupling is of electromagnetic strength.
Even much more weakly coupled leptoquarks can be observable, especially in
the $e\gamma$ mode. For all processes the largest cross-sections correspond
to leptoquarks of charge $-$5/3 and $-$1/3. Finally, in $\gamma\gamma$
collisions there is the possibility of singly producing second and third
generation leptoquarks although there are suppression factors relative to
the first generation.

\begin{figure}
\vspace{3.cm}
\caption{Diagram giving the dominant contribution to $\gamma
\gamma\rightarrow S e\bar q$}
\end{figure}

\vskip.5cm

\bibliographystyle{unsrt}

\end{document}